\documentclass[twocolumn,prb]{revtex4}
\usepackage{amsmath,graphicx,color}
\begin{document}
\author{Trithep Devakul$^1$, Satya N. Majumdar$^2$ and David A. Huse$^1$
}
\affiliation{
    $^1$ Department of Physics, Princeton University, NJ 08544, USA\\
    $^2$ Laboratoire de Physique Th´eorique et Mod\'eles Statistiques, Universit\'e Paris-Sud, Orsay, France.
}
\title{Probability distribution of the entanglement across a cut
at an infinite-randomness fixed point}
\begin{abstract} 
    We calculate the probability distribution of entanglement entropy $S$ across a cut of a finite one-dimensional spin chain of length $L$ at an infinite-randomness fixed point using Fisher's strong randomness renormalization group (RG).  
    Using the random transverse-field Ising model as an example, the distribution is shown to take the form $p(S | L) \sim L^{-\psi(k)}$, where $k\equiv S/\log\left[L/L_0\right]$, the large deviation function $\psi(k)$ is found explicitly, and $L_0$ is a nonuniversal microscopic length.
    We discuss the implications of such a distribution on numerical techniques that rely on entanglement, such as matrix product state (MPS) based techniques.
    Our results are verified with numerical RG simulations, as well as the actual entanglement entropy distribution for the random transverse-field Ising model which we calculate for large $L$ via a mapping to Majorana fermions.
\end{abstract}
\maketitle
\section{Introduction}

Entanglement has emerged as a key ingredient in the study of quantum many-body systems~\cite{}.
In particular, the matrix product state (MPS) or density matrix renormalization group (DMRG) methods for numerically studying quantum states work directly with the bipartite entanglement of a pure many-body quantum state across each ``cut''.  In a disordered system, this entanglement may vary widely between different locations of the cut.  Thus it is of interest to understand the probability distribution of the entanglement for such states.  We are particularly interested here in highly-excited eigenstates that are of interest in studies of many-body localization (MBL).

At quantum criticality, the ground-state entanglement of certain translationally-invariant systems is known to display universal behavior characteristic of the associated conformal field theory~\cite{}.
In one spatial dimension, the von Neumann entropy (in bits) for a subsystem of length $L$ at such a conformally-invariant quantum critical point
scales for large $L$ as \cite{vidal}
\begin{equation}
    S = -\text{Tr}\{\rho \log_{2} \rho\} = \frac{c}{3}\log_{2} L + \text{const} ~,
    \label{eq:purescaling}
\end{equation}
where $c$ is the central charge~\cite{} and $\rho$ is the reduced density operator for the subsystem.

A similar scaling of entanglement as the logarithm of $L$ applies for a number of disordered one-dimensional systems whose ground states are governed by infinite-randomness fixed points of strong-randomness renormalization groups\cite{refael-moore}. 
This class of systems include the disordered Heisenberg and XXZ spin chains, and the transverse field Ising model (TFIM)\cite{fisher1,fisher2}.  For the disordered TFIM, as long as it can be transformed to noninteracting fermions, this scaling of the entanglement also applies to all excited eigenstates\cite{rsrgx}.

With the surge of interest in the many body localization (MBL) phase transition, which might be described by an infinite randomness fixed point \cite{VHA, PVP, liangsheng}, a deeper understanding of this entanglement scaling becomes even more relevant.  Also, critical points within the MBL phase, should they exist, are governed in one-dimensional systems by infinite-randomness fixed points and exhibit this logarithmic scaling of the entanglement (Eq~\ref{eq:purescaling}) with an effective (irrational) central charge\cite{rsrgx,VA, VPP}.

The practical importance of quantum entanglement is highlighted by the success of the density matrix renormalization group (DMRG) algorithm, which relies on the low entanglement nature of certain quantum states to accurately represent them as matrix product states (MPS)~\cite{schollwock}.
There has been progress in obtaining highly excited eigenstates of MBL systems using this MPS framework~\cite{pc1,pc2,serbyn}.
A variant of DMRG called DMRG-X has been developed to treat highly-excited states, with the specific goal of treating MBL systems~\cite{dmrgx}.
The accuracy of these algorithms depends on the bond dimension $\chi$ allowed on each of the internal bonds of the MPS.
Representing a state to a certain accuracy requires the underlying MPS to have a bond dimension $\chi$ on a bond which will generally scale exponentially with the entanglement entropy $\chi \sim e^{aS}$, where $a$ depends on the structure of the entanglement spectrum (eigenvalues of the reduced density matrix).
The logarithmic growth of $S$ with $L$ thus leads to a polynomial growth in the necessary $\chi$ and therefore of the computational complexity of the DMRG.
In a disordered system, $S$ and thus the necessary bond dimension $\chi$ varies along the spin chain, so to evaluate the scaling of the DMRG algorithm's computation time we need to understand the distribution of the entanglement.

In this paper, we derive an explicit expression for the probability distribution of the entanglement entropy at the infinite randomness fixed point of the critical random TFIM for a finite open system, like those studied by DMRG\@.  From the perspective of Fisher's strong disorder renormalization group (RG) for the random critical TFIM\cite{fisher1}, there is no difference between a ground or excited state\cite{rsrgx},
and so our results may also be applied to DMRG-X calculations.
This expression is used to derive the scaling behavior and probability of finding entanglements from various points in the distribution.  
A finding is that the computational time is not dominated at large sample length $L$ by the local maximum of the entanglement.  We check our findings with numerical simulations of the RG itself, as well as calculations on the disordered transverse-field Ising model, and find excellent agreement.

\section{Fisher's renormalization group}

We first begin with a rough overview of 
Fisher's RG analysis~\cite{fisher1,fisher2} for the random transverse-field Ising chain Hamiltonian of the form $\mathcal{H} = \sum_i J_i \sigma_i^x \sigma_{i+1}^x - \sum_i h_i \sigma_i^z$.
At criticality,  $J_i$ and $h_i$ are independent random couplings drawn from the same distribution (which, for convenience, we assume has equal weights for positive and negative couplings).
By performing a Jordan-Wigner transformation in the $\sigma_z$ basis, this Hamiltonian can be expressed 
in terms of the conventional Majorana fermions $\gamma_j$, 
with only nearest neighbor couplings:  
$\mathcal{H} = i \sum_{j=1}^{2L-1} g_j \gamma_j \gamma_{j+1}$,
with $g_{2i-1} = h_i$ and $g_{2i} = J_i$.

The RG proceeds by always treating
the strongest coupling, in exactly the same way as is done for a random singlet phase.  
The energy scale $\Omega$ is defined to be the strongest of all the couplings $\Omega=\max_j |g_j|$, beginning at $\Omega=\Omega_0$ for the unrenormalized bare model.
An RG ``step'' begins by finding the coupling with $|g_j| = \Omega$.
Since the distribution of the $g$'s is very broad, we will almost always have $|g_{j-1}|\ll|g_j|\gg|g_{j+1}|$.
As a result, the two Majoranas $\gamma_j$ and $\gamma_{j+1}$ form a two-level system with eigenenergies $\pm g_j$ that is only weakly coupled to neighboring Majoranas.  We put this two-level system in one of its local eigenstates (Fisher always chose the ground state, since that is what he was focused on) and then treat the coupling to its neighbors perturbatively.
This results, from leading order in degenerate perturbation theory, in an effective coupling $g' = \pm g_{i-1} g_{i+1} / g_i$ between the neighboring Majoranas $\gamma_{j-1}$ and $\gamma_{j+2}$, with the sign of the new coupling depending on whether the ground or excited state was chosen.
After each such decimation, the energy scale $\Omega$ is decreased accordingly.  Note that the entanglement structure and the magnitudes of the renormalized couplings are independent of whether the ground or excited state was chosen.  At this quantum critical point, this RG flows to ``infinite randomness'', meaning the probability distribution of $\log{|g|}$ becomes arbitrarily broad, and the approximation of keeping only the leading-order perturbative coupling becomes asymptotically exact\cite{fisher1}.

It is convenient to use the scaled log couplings $\beta_i = \log (\Omega / g_i)\geq 0$ and log energy cutoff $\Gamma = \log (\Omega_0 / \Omega)$, which is the RG flow parameter.
After a decimation, the new $\beta$ is then simply given by $\beta ' = \beta_{i-1} + \beta_{i+1}$, greatly simplifying the flow equations.
Solving the flow equations results in the fixed point distribution\cite{fisher2}
\begin{equation}
    p(\beta/\Gamma) = e^{-\beta/\Gamma}~.
    \label{eq:pbeta}
\end{equation}

\begin{figure}[t]
    \centering
    \def\svgwidth{0.45\textwidth}
    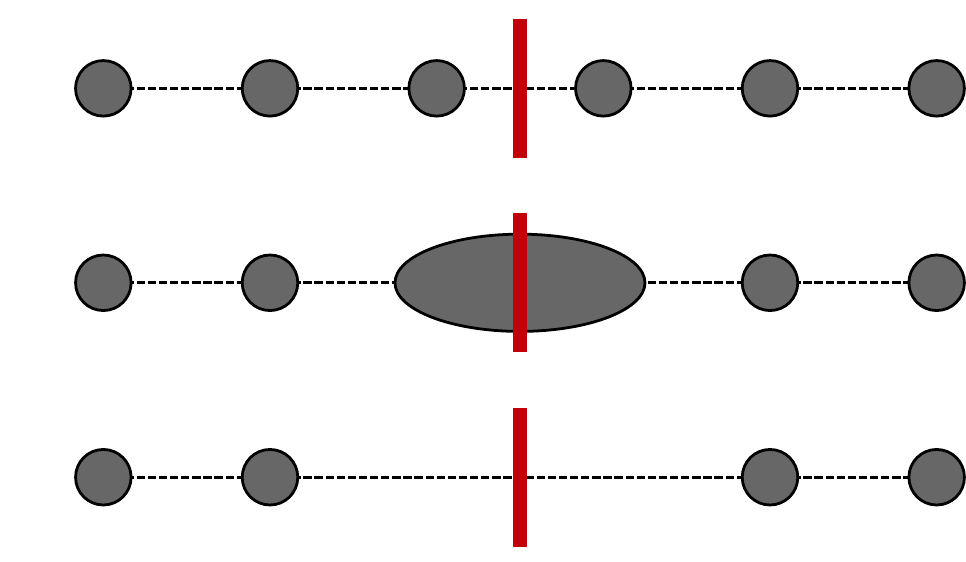%
    \caption{Illustration of how entanglement is created across a cut in the TFIM.  The bond crossing that cut must first be decimated ($1\rightarrow2$) resulting in the two spins on either side being fully correlated, but not yet entangled.  The cut now lies ``within'' that combined spin.  When the field on this site is decimated ($2\rightarrow3$), the state of that spin is put into an equal superposition of the two possible correlated states, thus becoming entangled.  The site is frozen out, there is now one new bit of entanglement across the cut, and the cut once again lies on a bond.  Thus, the amount of entanglement $S$ across the cut is half the number $N$ of decimations across it, $S = N/2$.}
    \label{fig:tfim}
\end{figure}

In the original spin language, a field $h_i$ being ``integrated out'' corresponds to the local state of the renormalized spin $\tilde\sigma_i^x$  being fixed to $\left|\rightarrow\right\rangle$ or $\left|\leftarrow\right\rangle$, which more microscopically represents an equal-amplitude linear combination of some particular pattern of the bare $\sigma_i^z$'s within the cluster and its opposite under flipping all spins.  This step thus introduces one bit of entanglement across any cut within this cluster.
A bond $J_i$ being ``integrated out'' corresponds to the two renormalized spins $\tilde\sigma_i^z$ and $\tilde\sigma_{i+1}^z$ being fixed to be either parallel or antiparallel and thus combined to be one renormalized spin, so it introduces new microscopic correlations but no new entanglement.

In this picture, an entanglement cut obtains
one new bit of entanglement whenever a field $h_i$ is decimated across the cut, which is precisely half of all decimations.  The decimations across the cut alternate between $h_i$'s and $J_i$'s:  When the cut is between two renormalized spins it is on a bond $J_i$.  When this $J_i$ is decimated these two spins are fully correlated and the cut is then within the new renormalized spin and thus ``on'' a field $h_i$.  When that $h_i$ is decimated the entanglement increases by one bit and the cut returns to being on a bond between two renormalized spins.  This process is shown in Fig~\ref{fig:tfim}.  Thus when the number of decimations across a cut is $N$, the entanglement across that cut is $S = N/2$ bits.

\section{Derivation of the entanglement distribution}
To obtain the probability distribution for the the number of decimations $N$ across a single cut for a system of fixed total length $L$, we first instead consider the problem of having a system at a fixed log energy cutoff $\Gamma$.
This is significantly simplified due to the fact that the ``steps'' in $\log \Gamma$ between two decimations are, at the fixed point, independent and identically distributed (i.i.d.) events~\cite{refael-moore}.  

Suppose on running this RG the most ``recent'' decimation across our cut was at log energy cutoff $\Gamma_0$.  
Once the cutoff has increased to a higher $\Gamma$, what is the probability $R(\Gamma,\Gamma_0)$ that no more decimations across our cut have happened?
At the critical fixed point, this probability depends \emph{only} on $\ell \equiv \log \Gamma / \Gamma_0$, 
so $R(\Gamma,\Gamma_0)=R(\ell)$.  
Solving the flow equations~\cite{refael-moore}, one finds 
\begin{eqnarray}
    R(\ell) 
    &=&  \left( \frac{3+\sqrt{5}}{2\sqrt{5}} e^{-\frac{3-\sqrt{5}}{2} \ell} - \frac{3-\sqrt{5}}{2\sqrt{5}} e^{-\frac{3+\sqrt{5}}{2} \ell} \right)~.
    \label{eqn:R(l)}
\end{eqnarray}
And therefore, the probability $r(\ell)$ that the next decimation occurs at $\ell$ is given by
\begin{equation}
    r(\ell) = -\frac{d R(\ell)}{d\ell} = \frac{1}{\sqrt{5}}\left( e^{-\frac{3-\sqrt{5}}{2} \ell} - e^{-\frac{3+\sqrt{5}}{2} \ell} \right)~.
    \label{eqn:r(l)}
\end{equation}

Now we may proceed towards the distribution of total decimations $N$ between $\Gamma_0$ and a final cutoff $\Gamma$. 
We begin with a decimation that occurred initially at $\Gamma_0$, followed by $N$ decimations happening at $\left\{  \Gamma_1, \Gamma_2, \dots, \Gamma_N \right\}$, with all $\Gamma_{i-1}<\Gamma_i < \Gamma$.
Taking advantage of the i.i.d.\ nature of the steps in $\log \Gamma$, we define $\ell_n = \log \Gamma_n / \Gamma_0$, and $\ell = \log \Gamma / \Gamma_0$, so that the probability of having $N$ decimations is obtained by integrating over all possible $\left\{ \ell_1,\dots,\ell_N \right\}$ with their respective probabilities,
\begin{equation}
    p(N | \ell) = \int \left[ \prod_{n=1}^N r(\ell_n-\ell_{n-1})d\ell_n \right] R(\ell-\ell_N)~.
    \label{}
\end{equation}

This integral can be expressed as a convolution in the variables $\Delta \ell_n = \ell_n - \ell_{n-1}$, which we can then apply the method of Laplace transforms to, resulting in
\begin{equation}
    \tilde{p}(N | a) = \int_0^{\infty} p(N | \ell) e^{-\ell a}d\ell = {\left[\tilde{r}(a)\right]}^N \tilde{R}(a)
    \label{}
\end{equation}
where $\tilde{\cdot}$ denotes the Laplace transformation, and $a$ is the Laplace conjugate variable.
$\tilde{r}(a)$ can be calculated explicity from Eq.~\ref{eqn:r(l)} to be $\tilde{r}(a)={(1+3a+a^2)}^{-1}$.

To invert the Laplace transform, we employ the Bromwitch integral
\begin{eqnarray}
    p(N|\ell) &=&  \frac{1}{2\pi i}\int_{\gamma - i\infty}^{\gamma + i\infty} da~e^{a \ell} \tilde{p}(N|a) \\ 
&=& \frac{1}{2\pi i}\int_{\gamma - i\infty}^{\gamma + i\infty} da~\tilde{R}(a)\exp \left[ \ell H(a;N/\ell)\right]
\end{eqnarray}
where 
\begin{equation}
    H(a;x) = a+x\log\tilde{r}(a)
    \label{}
\end{equation}
and $\gamma$ is chosen such that the contour of integration is to the right of all singularities of the integrand in the complex plane.
In the limit of large $\ell$, with $N/\ell$ kept finite, this integral can be well approximated by the saddle point method.
By analyticity of $H$, the saddle point for $x>0$ is on the real $a$ axis at the minimum of $H(a;x)$ to the right of both of its singularities. 
Then, up to polynomial prefactors, we have that 
\begin{equation}
    p(N|\ell) \sim \exp\left[-\ell \phi (N/\ell)\right]
    \label{eq:p_N_ell}
\end{equation}
where $\phi(x) = -\min_a H(a;x)$.

\begin{figure}[t]
    \centering
    \includegraphics[width=0.45\textwidth]{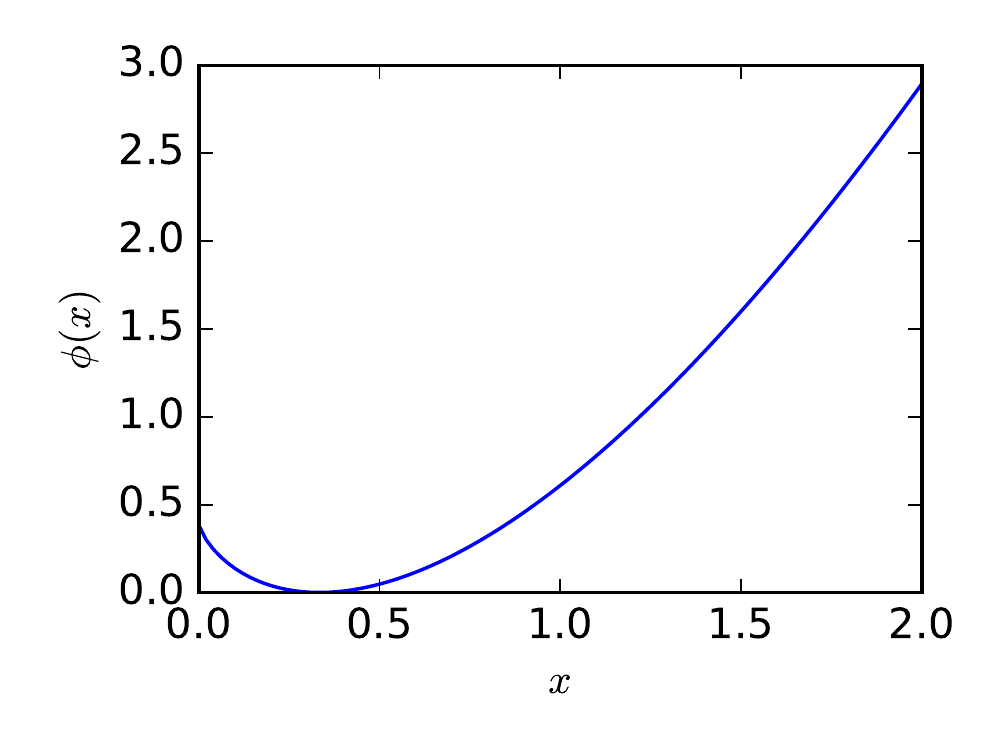}
    \caption{Plot of the function $\phi(x)$ (Eq.~\ref{eq:phi}).  }
    \label{fig:phix}
\end{figure}

Minimizing $H(a;x)$ we get
\begin{eqnarray}
    \phi(x) &=&  \frac{3-\sqrt{5+4{x}^2}-2x}{2}  + \nonumber \\
    && x\log\left[ 2{x}^2 + x\sqrt{5+4{x}^2} \right]~,
    \label{eq:phi}
\end{eqnarray}
which is shown in Fig~\ref{fig:phix}.
This function has a finite value at $x=0$ of $\phi(0) = (3-\sqrt{5})/2$, reaches a minimum at $\phi(1/3)=0$, and goes as $\phi(x) \approx 2x(\log(2x)-1)+3/2+\mathcal{O}(1/x)$ for large $x$.  

The connection to a system of fixed size can now be done by relating\cite{fisher1} $L = \Gamma^2$, up to some proportionality constant (which we set to $1$ for now as it does not affect any of our immediate conclusions).
This works because in the large $\ell$ limit, the fluctuations of log lengths are of order one: $\log L = 2\log \Gamma + \mathcal{O}(1)$.
Thus, after the substitution $\ell = \frac{1}{2}\log L$, for the critical random TFIM the distribution of the bipartite entanglement $S = N/2$ (in bits) across a cut takes on the form
\begin{equation}
    p(S | L) \sim L^{-\psi(k)}
    \label{}
\end{equation}
for a finite sample of length $L$ with open ends, with $\psi(k) = \frac{1}{2}\phi(4k)$ and $k \equiv S / \log L$.

There are a few interesting regions in this function that deserve mentioning.
The fraction of cuts with zero entanglement is non-zero but vanishingly small, scaling as $L^{-\theta_0}$, with $\theta_0= \psi(0)= \cong 0.191$.
Meanwhile, the typical (most likely) entanglement is at $k = 1/12 \cong 0.083$. 
This is in agreement with the mean entanglement entropy for the TFIM of $S \approx \frac{1}{12}\log L$ for a single cut~\cite{refael-moore,note}.
Also of interest is the typical largest entanglement cut to appear in a sample.
This is the $S$ which appears with probability scaling $p(S|L) \sim L^{-1}$, of which there are $\mathcal{O}(1)$ of in each sample.
This happens when $\psi(k)=1$, which is at $k \cong 0.417$, much larger than the typical $k$.

The cumulant generating function for this distribution $g(t|\ell) = \log \langle e^{Nt} \rangle$ has been obtained analytically previously~\cite{moore}.
To see the relation between these two results, one can express in the limit of large $\ell$ (using the same saddle point approximation),
\begin{eqnarray}
    g(t|\ell) &=& \log \int_0^{\infty} p(N|\ell) e^{Nt} dN\\
    &\approx& -\ell \min_x \left\{ \phi(x) - tx \right\} + \dots\\
    &=& {-\frac{3-\sqrt{5+4 e^t}}{2}\ell} + \dots
    \label{}
\end{eqnarray}
in agreement with Ref.~\onlinecite{moore} at large $\ell$ up to additive logarithmic corrections.  
Thus, the connection between $\phi(x)$ and $g(t|\ell)$ is via a Legendre transform.

\section{Implications for DMRG}
What does this distribution of entanglement mean for numerical techniques such as DMRG that rely on entanglement?
To accurately represent a state as a matrix product state (MPS), the number of states kept (or bond dimension) $\chi$ is related to the entanglement across a cut on that bond.
Having a high bond dimension allows DMRG to capture states more accurately, but at the cost of increased computation time.  
At the infinite randomness fixed point, the entanglement comes only in the form of $S$ maximally entangled pairs, so 
the entanglement spectrum is therefore simply $2^S$ equal nonzero eigenvalues, followed by zeros, for which a bond dimension of $\chi=2^S$ is needed.  
Using an algorithm that allots bond dimension independently for each bond as needed to attain a certain accuracy, we can ask the question of how the distribution of entanglement affects the scaling of the computation time for such a numerical technique.

The DMRG-X algorithm for finding highly excited MBL eigenstates~\cite{dmrgx} relies on the diagonalization of a $d^2\chi^2 \times d^2\chi^2$ effective Hamiltonian, where $d=2$ is the local physical number of degrees of freedom.
This diagonalization can be done in full, or using alternatives such as shift-invert Lanczos.  
The computational complexity of these algorithms scale as some power $c$ of the bond dimension.  For full exact diagonalization, $c=6$, while $c$ would be smaller for methods that scale more favorably at large bond dimension. 
Letting $T$ denote the computation time, we have $T \sim \chi^{c}\sim L^{ck\log 2}$, where we have defined $k \equiv S/\log L$.

We can now ask what the scaling of the mean computational time $\langle T(L) \rangle$ is with system size $L$.
The average computation time per bond scales as
\begin{eqnarray}
    \langle T(L) \rangle &\sim& \int_{0}^{\infty} p(S|L) L^{ck\log 2} dS~.
    \label{}
\end{eqnarray}
Up to logarithmic corrections, for large $L$
\begin{eqnarray}
    \langle T(L) \rangle &\sim& L^{\max_k \left[ck\log 2 - \psi(k)\right]}~.
    \label{}
\end{eqnarray}
For full exact diagonalization, $c=6$ and one finds the time to be dominated by bonds with $k^* = 2/\sqrt{37} \cong 0.329$, resulting in $\langle T(L) \rangle \sim L^{\alpha_0}$, where $\alpha_0= c k^*\log2 - \psi(k^*) \cong 0.771$.
The mean time for treating every bond in the full sample thus scales as $L^{1+\alpha_0}$.  These powers will only be smaller if a method that scales more favorably than full exact diagonalization can be used.
This ignores the strong correlation of entanglement entropy between bonds within a sample, which are important but don't affect these exponents for the \emph{mean} time.

The maximum entanglement bonds were found earlier to occur near $k\cong 0.417$, which are higher in entanglement and rarer than the ones that dominate the computation time.
This is important, as it means that most samples will have a few bonds of this high difficulty, but most of the computational time will still be spent on somewhat easier bonds that appear with higher frequency.
Indeed, this indicates that the computational time for a sample will generally not be determined by its most difficult bond, for which there may be large sample-to-sample variation; this suggests that the computation time should ``self-average''.

\section{Numerical results}
To check our analytic findings, we numerically run the infinite randomness RG on many samples.
The system is treated as an array of blocks which are the bonds between the Majorana modes in the renormalized model.  Each block $i$ has a coupling strength $g_i$, a length $l_i$ and a normalized distribution $p_i(N)$ for the number of decimations $N$ across all internal cuts within the block.  
The blocks are initialized with a coupling $g_i$ from the fixed point distribution (Eq~\ref{eq:pbeta}) with $\Gamma_0=1$, an initial length $l_i=1$, and a trivial initial internal distribution of $N$ of $p_i(N)=\delta_{N,0}$.
Upon decimation of a block $i$, the blocks $i-1$,$i$, and $i+1$ are merged to a single block, and the probability distribution is updated as
\begin{eqnarray}
    p'(N) = \frac{l_{i-1}p_{i-1}(N) + l_{i}p_{i}(N-1) + l_{i+1}p_{i+1}(N)}{l_{i-1}+l_{i}+l_{i+1}}~.
    \label{}
\end{eqnarray}
The coupling strength is updated as prescribed by the RG rules, and the new length is simply the summation.

To obtain the number of decimations given a fixed flow interval $\ell = \log \Gamma / \Gamma_0$, we simply run the RG until the log energy cutoff has reached the desired value.
Then, the probability distribution can be sampled from all the remaining blocks in the system weighted by their length.
The probability distribution for a system of fixed size $L$, on the other hand, can be obtained by initializing the system with an odd number of blocks $L$ (each of initial length $l=1$) and simply running the RG until only one block remains, which is guaranteed to have length $L$.

\begin{figure}
    \centering
    \includegraphics[width=0.45\textwidth]{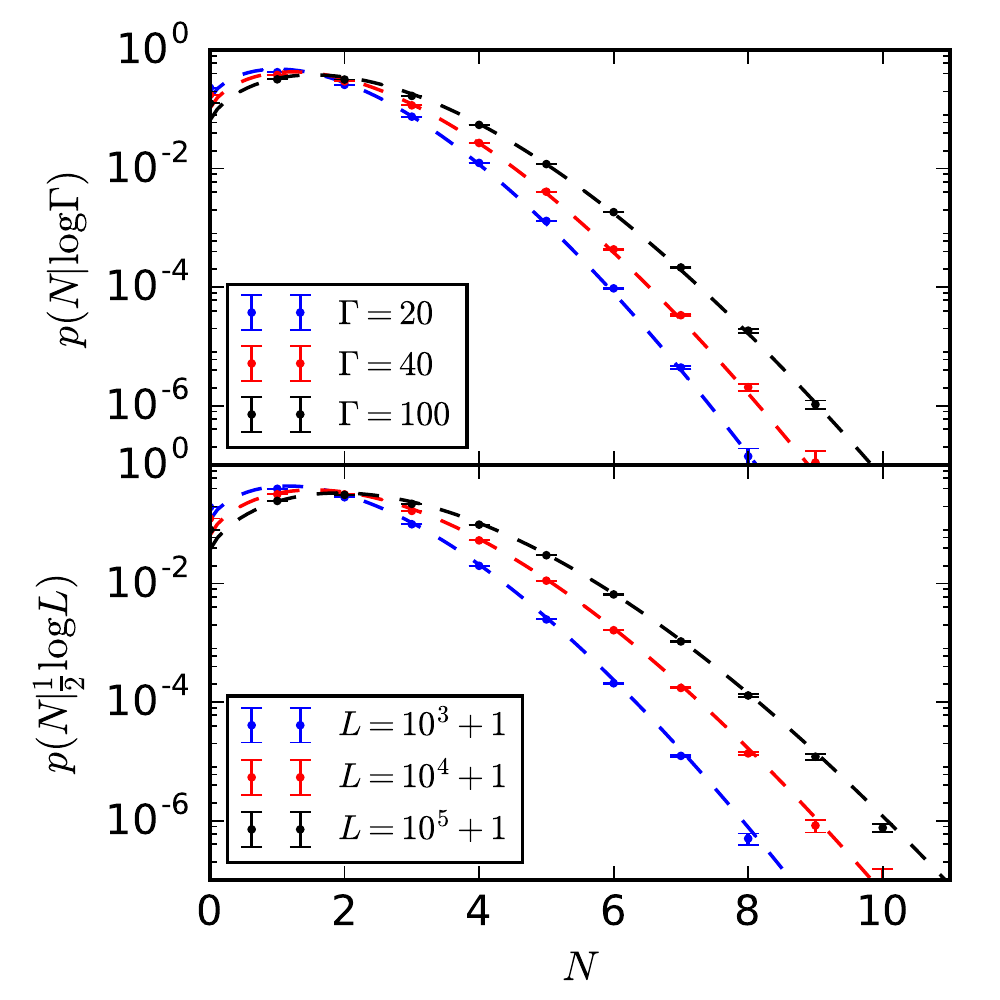}%
    \caption{   
        Probability distribution of $N$ from numerical RG simulations obtained by (top) running the RG until a fixed log energy cutoff $\Gamma$, and (bottom) by running the RG to completion for a fixed system size $L$.
        Dashed lines show the normalized analytical results (Eq~\ref{eq:p_N_ell}) with $\ell=\log \Gamma$ in the fixed $\Gamma$ case, and $\ell = \frac{1}{2}\log L$ in the case with fixed $L$.
        There is excellent agreement between the analytical and numerical results.
    }\label{fig:RandomSinglet}
\end{figure}

Figure~\ref{fig:RandomSinglet} shows the numerical results for fixed $\Gamma$, and fixed $L$, along with the (normalized) analytical predictions from Eq~\ref{eq:p_N_ell}. 
In the case of fixed $L$, the substitution $\ell = \frac{1}{2}\log \Gamma$ is used, taking advantage of the relationship between the mean $L$ and $\Gamma$.
In both cases, the agreement between numerical results and the approximate analytical expression is excellent.

\section{Transverse Field Ising Model}
Finally, we compare with results for the random transverse field Ising model, which can be obtained by a mapping to noninteracting majorana fermions and the entanglement entropy obtained from the single particle correlation matrix~\cite{peschel}.
Due to self duality, this system is critical when the distribution of $J$ and $h$ are the same.
We pick the disorder distribution to be the fixed point distribution Eq~\ref{eq:pbeta}.  
Since we can always rescale the energy, we set $\Omega=1$ so the cumulative distribution for $J$ is given by $P(J<j) = j^{1/\Gamma}$ for $0\leq j\leq 1$, and similarly for the on-site fields $h$. 
$\Gamma$ is the parameter that flows towards infinite randomness, so we expect the Fisher RG to hold for large $\Gamma$.  
At small $\Gamma$, the distribution becomes more and more narrow arround $J=h=1$, and so realistically our results cannot apply for very small $\Gamma$.
On the other hand, very high $\Gamma$ runs into problems with machine precision at large system sizes.
We will focus on $\Gamma=1$, which corresponds to a flat disorder distribution as is commonly used in studies of localization (cf. the Anderson model).

In comparing with actual microscopic models, there is a proportionality constant $L_0$, $\Gamma/\Gamma_0 = \sqrt{L/L_0}$, which depends on the microscopic details.
There is therefore a free parameter $L_0$ in defining $\ell = \frac{1}{2}\log L/L_0$.
However, 
\begin{equation}
\log \left[ p(S|\ell)\right]/\ell = -\phi(2S/\ell) + \text{const}(\ell)
\label{eq:collapse}
\end{equation}
is a universal function of $S/\ell$ for all $L$ (with the constant providing the normalization having some dependence on $L$).

\begin{figure}
    \centering
    \includegraphics[width=0.45\textwidth]{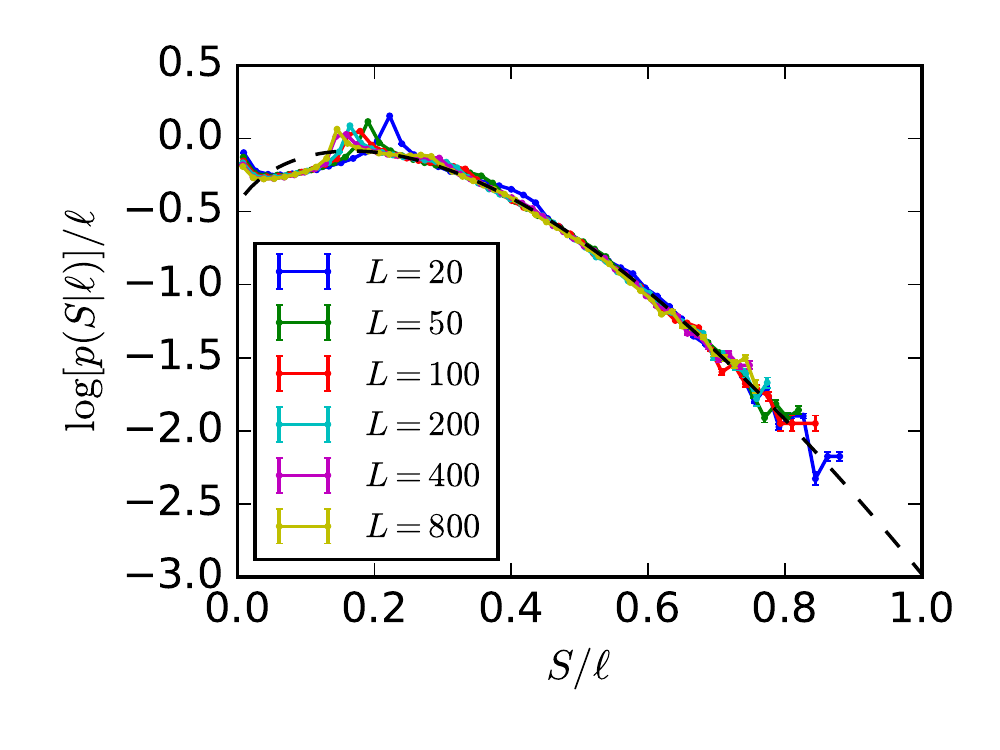}
    \caption{Distribution of entanglement entropy $S$ sampled from cuts of random excited eigenstates of the random transverse field Ising model of varying $L$.  
        The random fields and couplings are taken from the fixed point distribution (Eq~\ref{eq:pbeta}) with $\Gamma=1$, corresponding to a flat distribution.
        With $\ell = \frac{1}{2}\log L/L_0$ for $L_0 \approx 9.8\times 10^{-4}$, all these curves collapse well towards a single curve.
        The dashed line shows the analytic prediction (Eq~\ref{eq:collapse}) with a fixed constant shift to align with the numerical data.
    }\label{fig:Peschel}
\end{figure}

Figure~\ref{fig:Peschel} shows the entanglement distribution for the TFIM with $\Gamma=1$ in this manner.
For all $L$, the entanglement distribution shows a peak at an entanglement of $S = 1$ bit, and collapses very nicely to a single curve in the tail, in very good agreement with the analytical curve.
Notice that the fitted $L_0$ is typically much smaller than 1, this is a result of the fact that for $L \ll \xi$ the localization length, entanglement will grow much quicker than $\log L$, and so at large $L$, $S\sim \log L/L_0$ will appear to have a small $L_0$.

\section{Conclusion}
We have successfully obtained, using Fisher's strong randomness RG, an analytic expression for the probability distribution of entanglement entropy for a random TFIM chain at the infinite randomness fixed point.
The distribution is found to have a large deviation form $p(S|L) \sim L^{-\psi(S/\log [L/L_0])}$ with  $\psi(x)$ given explicitly.
Although the results were obtained for a system at a finite log energy cutoff $\Gamma$, they can equally be applied to the more applicable case of a finite length chain by the substitution $L \sim \Gamma^2$.
These results can also be applied to excited states as well, and are verified by numerical calculations.

The distribution of entanglement entropy is particularly relevant for DMRG studies.
The typical entanglement to appear in a sample grows logarithmically with $L$ and is in agreement with previous calculations for mean entanglement.
We find the typical maximum entanglement entropy to appear in a sample of length $L$, and discover that it is higher than the entanglement which dominates the computational time for DMRG using exact diagonalization of an effective Hamiltonian.

We thank Vedika Khemani and Shivaji Sondhi for stimulating discussions.


\begin{thebibliography}{2}
\bibitem{note}
    We only consider a finite system with one cut, while Refael and Moore\cite{refael-moore} consider a finite subsystem of an infinite system and thus two cuts, so our results differ by a factor of 2.
\end{thebibliography}
\end{document}